\documentclass[journal]{IEEEtran}

\ifCLASSINFOpdf
\else
  \usepackage[dvips]{graphicx}
\fi

\usepackage{url}
\usepackage{color, soul}
\usepackage{xcolor}
\usepackage{cite}
\usepackage[colorlinks,linkcolor=black, citecolor=black]{hyperref}
\usepackage{multirow}
\usepackage{amsmath}
\usepackage{bm}
\usepackage{graphicx}
\usepackage{comment}
\usepackage{longtable}
\usepackage{xparse}
\usepackage{booktabs}
\usepackage{subfigure}
\usepackage{caption}
 
\usepackage{array}
\usepackage{ulem}
\usepackage{amsmath}
\usepackage{amssymb}
\usepackage[ruled,vlined]{algorithm2e}
\usepackage{setspace}
\usepackage{colortbl}

\begin{document}

\title{Controllable Accented Text-to-Speech Synthesis}
 
\author{Rui Liu, \IEEEmembership{Member, IEEE},  Berrak Sisman, \IEEEmembership{Member, IEEE}, Guanglai Gao, Haizhou Li, \IEEEmembership{Fellow, IEEE}

% \vspace{-2mm}
\thanks{This paper was submitted on 22 Spe 2022 for review. The research by Rui Liu was funded by the High-level Talents Introduction Project of Inner Mongolia University (No. 10000-22311201/002)
% \textcolor{black}{(with Rui Liu as the Principal Investigator)}
and the Young Scientists Fund of the National Natural Science Foundation of China (No. 62206136). The work by Haizhou Li was partly supported by the Agency of Science, Technology and Research, Singapore, through the National Robotics Program under Grant No. 192 25 00054, and by Programmatic Grant No. A18A2b0046 from the Singapore Government’s Research, Innovation and Enterprise 2020 plan (Advanced Manufacturing and Engineering domain). (Corresponding author: Guanglai Gao)}

\thanks{Rui Liu is with the Department of Computer Science, Inner Mongolia University, Hohhot 010021, China. 
% He is also with the Department of Electrical and Computer Engineering, National University of Singapore, and Singapore University of Technology and Design (SUTD).
(e-mail: liurui\_imu@163.com).}
\thanks{Berrak Sisman is with the Department of Electrical and Computer Engineering at University of Texas at Dallas, United States. }
\thanks{Guanglai Gao is with the Department of Computer Science, Inner Mongolia University, Hohhot 010021, China.
(e-mail: csggl@imu.edu.cn)}
\thanks{Haizhou Li is with School of Data Science, The Chinese University of Hong Kong, Shenzhen 518172, China. He is also with University of Bremen, Faculty 3 Computer Science / Mathematics, Enrique-Schmidt-Str. 5
Cartesium, 28359 Bremen, Germany (e-mail: haizhouli@cuhk.edu.cn). }}

\markboth{PREPRINT MANUSCRIPT OF IEEE/ACM TRANSACTIONS ON AUDIO, SPEECH, AND LANGUAGE PROCESSING}%
{Shell \MakeLowercase{\textit{et al.}}: Bare Demo of IEEEtran.cls for IEEE Journals}

\maketitle

\begin{abstract}
Accented text-to-speech (TTS) synthesis seeks to generate speech with an accent (L2) as a variant of the standard version (L1). Accented TTS synthesis is challenging as L2 is different from L1 in both in terms of phonetic rendering and prosody pattern. Furthermore, there is no easy solution to the control of the accent intensity in an utterance. In this work, we propose a  neural TTS architecture, that allows us to control the accent and its intensity during inference. This is achieved through three novel mechanisms, 1) an accent variance adaptor to model the complex accent variance with three prosody controlling factors, namely pitch, energy and duration; 2) an  accent intensity modeling strategy to quantify the accent intensity; 3) a consistency constraint module to encourage the TTS system to render the expected accent intensity at a fine level. Experiments show that the proposed system attains superior performance to the baseline models in terms of accent rendering and intensity control. To our best knowledge, this is the first study of accented TTS synthesis with explicit intensity control.
\end{abstract}


\IEEEpeerreviewmaketitle

% \vspace{-3mm}
\section{Introduction}
\IEEEPARstart{A}{ccented} text-to-speech (TTS) synthesis aims to synthesis speech with foreign accent instead of native speech. In other words, it seeks to generate speech with an accent (L2) as a variant of the standard version (L1). Recently, much progress has been made in the quality of neural TTS synthesis for L1  speech~\cite{ren2020fastspeech,liu2020expressive}, for example, WaveNet \cite{vanwavenet}, Deep-Voice 1, 2, 3 \cite{arik2017deep,gibiansky2017deep,ping2018deep}, Char2Wav \cite{sotelo2017char2wav}, Tacotron 1, 2 \cite{wang2017tacotron,shen2018natural}, Transformer-TTS \cite{li2019transformerTTS} and FastSpeech1 \cite{ren2019fastspeech}.
Neural TTS has outperformed conventional concatenative and statistical parametric approaches, where neural vocoders play an important role~ \cite{yamamoto2020parallel,kong2020diffwave,kong2020hifi,ping2020waveflow}.
% Besides, FastSpeech2 \cite{peng2019parallel} extends FastSpeech 1 \cite{ren2019fastspeech} by using additional acoustic features such as log-scale pitch and energy so that they show more expressive speech generation.
\textcolor{black}{
Accent is characterized by a distinctive manner of expression  that is influenced by the mother tongue, social group of speakers, or spoken in a particular region~\cite{loots2011automatic}. 
Generally, people find it easier to speak with others within their own accent group. 
%It may affect human-robot commutation due to these accent-related perceptions.
Therefore, the wide adoption of speech applications, such as chatbot and movie dubbing, calls for the study of accented TTS synthesis.}

% Neural TTS models have shown a rapid progress,  These
% models mostly resort to autoregressive generation of acoustic feature, which suffers from slow inference speed and a lack of robustness (word missing and skipping). Recently, several works such as Paranet \cite{peng2019parallel} and FastSpeech 1 \cite{ren2019fastspeech} have proposed nonautoregressive TTS models to handle such issues and achieve fast inference speed and improved robustness over autoregressive models. 

% The majority of the TTS models aim to synthesize high-quality speech of a single speaker/language from the given text and have been extended to support multi speakers/languages.
%Despite the progress, there is an increasing demand for accented L2 speech generation, which requires TTS models to generate high-quality accented L2 speech that well captures the complex and dramatic accent variation information. It is valuable to improve the generation capability of TTS models on various L2 accents.

\textcolor{black}{
% The accent is a distinctive way of pronouncing a language, especially one associated with a particular country, area, or social class.
Phoneme substitution and accent variance are two key aspects in  accent rendering \cite{loots2011automatic,deri2016grapheme}.
Phoneme substitution refers to the situation where L2 speakers pronounce L1 phonemes differently~\cite{kolluru2014generating}.
On the other hand, L2 speakers exhibit complex prosody in terms of pitch, energy, and duration~\cite{Laura05accent}, that is referred to as accent variance. This paper is focused on the study of accent variance. }  

An L2-accented TTS system is expected to produce human-like natural speech, just like any TTS system, at the same time  to generate speech with an intended accent at an appropriate intensity \cite{Laura05accent}.
The state-of-the-art neural TTS systems typically are trained for a standard voice. It remains a challenge how to effectively control the prosody of accent when rendering speech. We identify two challenges when dealing with accent variance.

First,  inadequate modeling of accent variance often leads to a flat L2 speech without noticeable accent signature.
Note that an L2 speech is highly expressive particularly in prosody, such as pitch, energy and rhythm. In the statistical approach to TTS, such as hidden Markov model (HMM), there have been attempts to address this problem \cite{malatji2016creating,sefara2019hmm}.  In the neural TTS approach, FastSpeech2 \cite{ren2020fastspeech}, FastPitch \cite{lancucki2021fastpitch}, Meta-StyleSpeech \cite{min2021meta}, and DAFT-EXPRT \cite{zaidi2021daft},  pitch and energy predictors are studied.
It remains as a challenge how we model the accent variance effectively with interpretable prosodic attributes such as pitch, energy, and duration. To this end, we study a novel neural approach to model the accent variance.

Second, it is difficult to control the intensity of accent because of the lack of well-defined intensity descriptors. In the HMM-based TTS framework, Wutiwiwatchai et al. \cite{wutiwiwatchai2011accent} proposed an accent level adjustment mechanism for bilingual TTS synthesis, where %, which relies on cross-lingual phone alignment, HMM state mapping and HMM interpolation.
the accent level is adjusted by means of interpolation between HMMs of native phones and HMMs of corresponding foreign phones.
In a recent multilingual TTS study \cite{zhang2019learning},
% a novel multilingual TTS model is proposed.
% that employs the domain adversarial training to disentangle the accent identity from the speaker identity.
% For accent control, 
the accent level can be controlled by varying the domain adversarial weight~\cite{zhang2019learning}.
In another attempt \cite{liu2020multi}, accent is manipulated via tone or stress embedding input. Accent as perceived in human speech is subtle and at a fine level. In the prior attempts to control the accent intensity, there is no direct and measurable  correlation between the controlling factor and the natural accent intensity. The question is how to characterize the accent intensity, and employ the intensity to control the synthesis of L2 speech, which is another focus of this paper. 
%This paper addresses the accent intensity quantification problem.
% All the above works controlled the accent of speech by using some discrete (adversarial weight in \cite{zhang2019learning} or tone/stress embedding in \cite{liu2020multi}) or meaningless (adversarial weight in \cite{zhang2019learning}) signs.
% Therefore, we desire an accent TTS to deliver accent speech with preferred accent intensity from light to strong explicitly, directly and smoothly.

In this paper, we propose a neural TTS solution with controllable accent and its intensity, that is referred to as the CAI-TTS system. We address the above two challenges by studying three novel mechanisms. 1) An accent variation adaptor that models the complex accent variance. The accent variance adaptor seeks to project the L2 speaker identity (ID), L2 accent identity and its intensity into the output speech at a phoneme level.  2) An accent intensity modeling scheme that uses relative attribute \cite{parikh2011relative} technique to automatically learn the relationship between L1 and L2 speech, thus quantifying the accent intensity for L2 speech. 3) A consistency constraint module that ensures the synthesized L2 speech manifests the expected accent intensity precisely.

During run-time inference, the CAI-TTS system takes the L2 speaker ID, L2 ID and its intensity along with the phoneme sequence as input, and outputs the L2 speech with the expected accent intensity, thereby regulating the accented speech generation explicitly. The significant contributions of this work include,
 
\begin{itemize}
    \item We introduce a novel accented TTS synthesis paradigm that explicitly controls the accent and its intensity in output speech.  
 
    \item We successfully design and implement a neural architecture, CAI-TTS, with three novel mechanisms.
 
    \item  We show that the proposed CAI-TTS framework outperforms the baseline models and generates high-quality L2-accented speech.
 
\end{itemize}

The rest of this paper is organized as follows.  In Section \ref{sec:back}, we discuss the background to motivate our work. In Section \ref{sec:proposed}, we propose the CAI-TTS framework. 
We report the experimental results in Section \ref{sec:exp}. Finally, Section \ref{sec:con} concludes the study.

\section{Related Work}
\label{sec:back}
 
\subsection{Accent Variance Learning}
 
Recently, much progress was made in neural TTS for L1 speech \cite{wang2017tacotron,shen2018natural,li2019transformerTTS,ren2019fastspeech,li2020robutrans,lee2021multi}.
However, the state-of-the-art models are not designed to deal with accent variance. In speech analysis, there has been study on prosodic variance of speech in terms of pitch and energy \cite{eyben2010opensmile}. Pitch is a key feature that characterizes speech variance, and has a great impact on perceptual quality of speech~\cite{lancucki2021fastpitch}.
Energy indicates frame-level magnitude of mel-spectrum and directly affects the loss computed on mel-spectrum.
There have been valuable attempts towards learning the variance information.
For example, in FastSpeech2 \cite{ren2020fastspeech}, a variance adaptor is introduced, which has a pitch and an energy predictor, to learn the speech variance at frame-level.
However, L2 speech is far more complex in terms of speaking style and variation than L1 speech, that calls for further study of two research problems. % The L1 variance adaptor is not enough to learn the complex variance information for accented L2 speech, and there are two problems that need to be solved.

The first problem is how to represent an accent. The L1 variance adaptor is not explicitly trained to encode accent information.
% \textcolor{black}{Dear.Prof.Li, maybe The L1 variance adaptor is \textbf{not} explicitly...?}  
% guided by any accent-related information, while information such as L2 speaker and L2 accent etc. is crucial for modeling the complex accent variance.
% The second problem is the characterization of accent variance. It is noted that phoneme-level acoustic features  carry rich and long-term accent variance more than frame-level features~ \cite{angkititrakul2006advances,du2021diverse}.
% In addition, there could be information loss arising from log-scaling of acoustic features after a series of mathematical transformation. We assume that such information loss could be avoided  if we predict real values at linear scale directly.
% Addressing the problems, we propose a novel accent variance adaptor, which encodes  L2 speaker, L2 accent, and its intensity, and predict phoneme-level pitch, energy and duration at linear scale.
The second problem is the characterization of accent variance. It is believed that phoneme-level acoustic features  are more accent informing than frame-level features~ \cite{angkititrakul2006advances,du2021diverse,lancucki2021fastpitch}.
\textcolor{black}{There have been studies to model the prosodic variance for phonemes, such as FastPitch \cite{lancucki2021fastpitch}. However, FastPitch doesn't model the energy.}
Addressing the above two problems, we propose a novel accent variance adaptor, which encodes  L2 speaker, L2 accent, and its intensity, and predicts phoneme-level pitch, energy and duration.

% .-------yes, Prof.Li, FastPitch also use phoneme pitch, but no energy information. ---Actually I also afraid that the phoneme pitch and energy predictors are not very novel, while the accent variance adaptor is novel.}

% , while Meta-StyleSpeech \cite{min2021meta} and DAFT-EXPRT \cite{zaidi2021daft} adopt the log-scale values for pitch and energy.

%In this paper, we propose to adopt a learnable attribute ranking function to quantify the accent intensity. 

\begin{figure*}[t]
    \centering
    % \vspace{-4mm}
    \centerline{\includegraphics[width=0.87\linewidth]{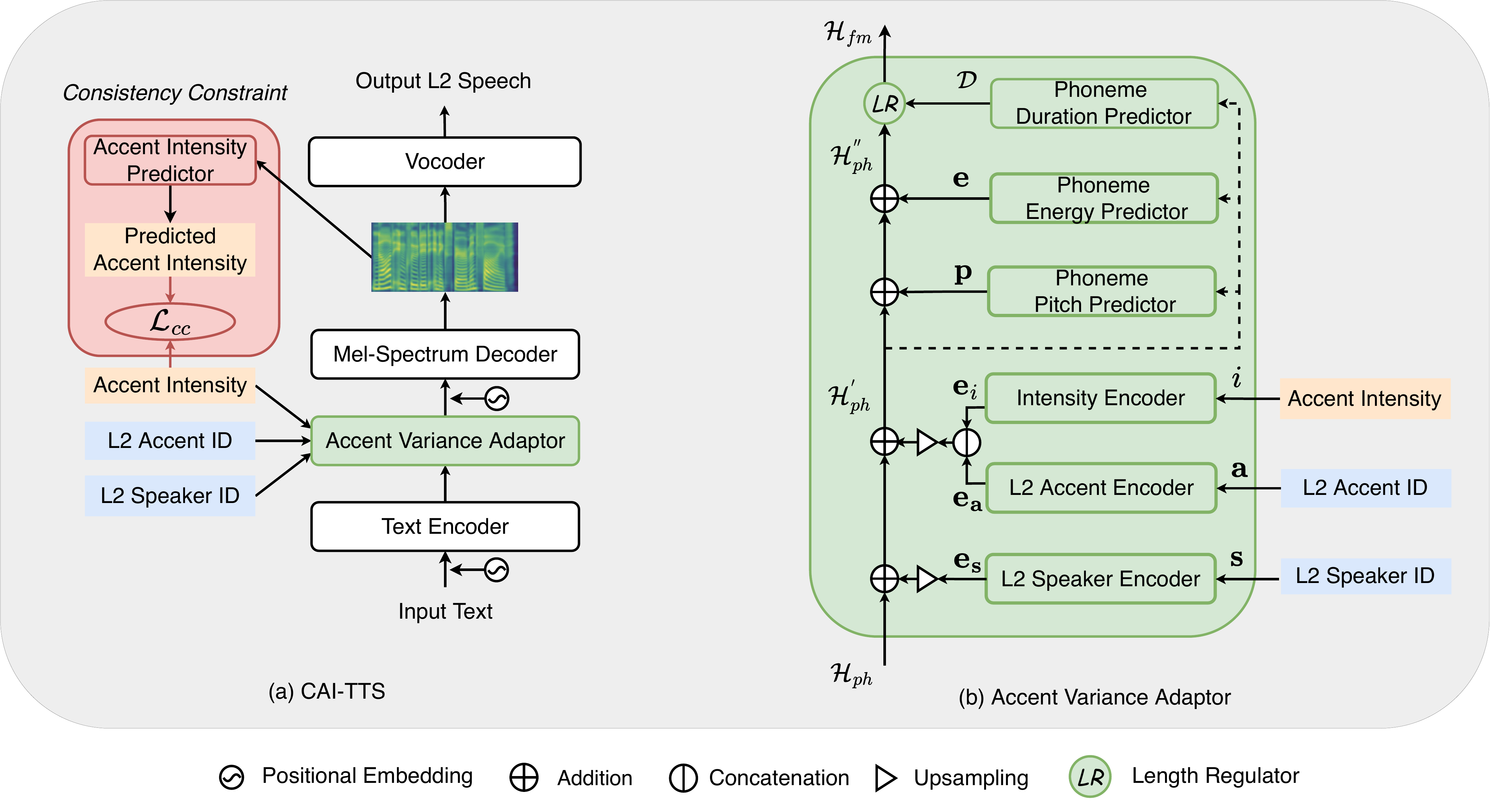}}
    % \vspace{-3mm}
    \caption{The diagrams for the proposed CAI-TTS system. (a) shows the overall model architecture; (b) shows the components of an accent variance adaptor, which includes encoders to encode the controlling factors, and predictors to generate accented prosodic embeddings and phoneme duration.}
    %(c) illustrates an accent intensity learning workflow.%}
    \label{fig:overall}
    % \vspace{-5mm}
\end{figure*}

\subsection{Attribute Ranking}
\label{subsec:rank}
 
Attribute ranking learns the difference between two samples that are significantly different in a particular attribute. It has been widely studied in computer vision~\cite{parikh2011relative,siddiquie2011image,meng2018efficient,saquil2020rank3dgan}.

Parikh et al. \cite{parikh2011relative} first proposed the concept of ``relative attributes'' in which pairs of visual entities can be compared, with respect to their relative strength of any specific attribute.
With a set of paired samples, they learn a global ranking function for each attribute that can be used to compare a pair of new samples respective to the target attribute.
Bringing this idea forward, we propose to compare the acoustic features from a pair, {$<$L1, L2$>$, of speech samples of the same speaker} and determine the difference between them to quantify the accent intensity.

% Some recent works \cite{zhu2019asru,lei2021fine} in speech processing also adopts similar idea, which totally different from our work.
% Note that they focus on emotion variance while our work focus on accent variance. Emotion changes in speech and accent changes have completely different expressions.

In emotional TTS synthesis~\cite{zhu2019asru,lei2021fine}, the idea of relative attributes has been explored. However,  accented TTS presents us a new challenge. In emotional TTS, the $<$neutral, emotional$>$ paired speech is available for emotion strength learning. However, the reference L1 speech sample is not available for accent intensity modeling. Learning of ranking functions is not possible without parallel data. Furthermore, the specific acoustic representations of emotion and accent attributes are completely different, that calls for a new solution to the problem. %. Therefore, the quantification of accent attributes cannot be solved using the same analysis mode.

In this work, we start with a L2 speech corpus, and construct a $<$L1, L2$>$ paired speech corpus by synthesizing high-quality L1 speech for each L2 speaker. We introduce a ranking function learning method to model the accent intensity from the $<$L1, L2$>$ corpus.

\subsection{Controllable Expressive TTS}
\label{subsec:pctts}

% \hl{xxxxxx}

% Neural TTS systems are now increasingly popular, improving upon older systems \cite{tokuda2002hmm} in synthesized
% speech naturalness. Non-autoregressive models \cite{ren2019fastspeech,ren2020fastspeech} are faster at inference than autoregressive models with about comparable naturalness of speech quality.
% However, going beyond the naturalness of speech, there has been considerable effort to improve the expressiveness of the renderings.
% Non-autoregressive models \cite{ren2019fastspeech,ren2020fastspeech} are faster at inference than autoregressive models with about comparable naturalness of speech quality.
%Following the enormous success of neural TTS, the focus of research in recent years is shifted to controllable expressiveness modeling. Specifically, 
There are two general approaches to sythesize expressive speech. 
1) Global speaking style: The speaking style can be obtained from a reference utterance, or a manual setting. Wang et al. \cite{wang2018style} adopt a reference encoder to derive a style embedding from an input utterance as a linear combination of some basis style vectors. Others \cite{hsu2018hierarchical,zhang2019learning,sun2020fully,liu2022controllable} use the variational autoencoder (VAE) to model latent representations for styles and prosody of speech. 
2) Token-wise fine-grained prosody embeddings: In FastSpeech2 \cite{ren2020fastspeech}, a variance adaptor is studied to add the variance information (e.g., duration, pitch, energy, etc.) to the phoneme hidden sequence in a way to generate expressive speech. 
Furthermore, more variance information has been added in the variance adaptor, such as emotion \cite{DBLP:conf/interspeech/SivaprasadKG21,im2022emoq,zhang2022iemotts} and style \cite{zaidi2021daft,lee2021styler,raitio2022hierarchical} to control the fine-grained expressiveness accordingly.

Strictly speaking, the above techniques don't really control the accent rendering. At run-time inference, the system generates the output speech by conditioning on the style embedding or the variance model. In this way, the speech is synthesized to follow a given style or prosody pattern. There is a lack of direct control over the synthesized speech, and there is no measurable  correlation between the controlling factor and the intended accent intensity.

To the best of our knowledge, there has no intuitive method for controlling the fine-grained accent in the literature. Motivated by the previous findings, we propose a novel neural architecture that quantitatively controls the accent and its intensity for the first time.

\section{CAI-TTS Neural Architecture}
\label{sec:proposed}
 
%We now describe the overall CAI-TTS model architecture, and the accent intensity learning scheme.

We propose a neural architecture, termed as CAI-TTS, as shown in Fig. \ref{fig:overall} (a) that consists of a text encoder, an accent variance adaptor, a mel-spectrum decoder, an additional accent intensity predictor and an universal HiFi-GAN vocoder.  The text encoder encodes the input phoneme sequence into phoneme embedding. The accent variance adaptor modulates the input phoneme embeddings towards the target accent. The mel-spectrum decoder converts the modulated phoneme embeddings into a mel-spectrum sequence.
Note that the accent intensity of the output L2 speech is estimated from the generated mel-spectrum sequence by the additional accent intensity predictor during CAI-TTS training. We impose a consistency constraint loss to minimize the difference between the estimated accent intensity and the expected intensity.  
Finally, the universal HiFi-GAN vocoder was used to synthesize high-quality L2 speech. 

The text encoder and mel-spectrum decoder share a similar architecture with FastSpeech2 \cite{ren2020fastspeech}. We use the feed-forward Transformer block, which is a stack of self-attention \cite{vaswani2017attention} layer and 1D-convolution, as the basic structure.
%In the following subsections, we describe the detailed design of accent variance adapter and consistency constraint for accent intensity in our CAI-TTS.
We denote the text encoder as $E_{t}(\cdot)$, which produces the phoneme hidden representation $\mathcal{H}_{ph}=E_{t}(\mathcal{P}+P\!E)$,
% \vspace{-2mm}
% \begin{equation}
%     \mathcal{H}_{ph}=E_{t}(\mathcal{P}+P\!E)
% \vspace{-2mm}
% \end{equation}
where $\mathcal{P}$ is the phoneme sequence, and $P\!E$ is a triangle positional embedding \cite{vaswani2017attention} to indicate the positional information.

In Sec. \ref{subsec:ava}, we will introduce the details of the accent variance adaptor and the consistency constraint for accent intensity predictor. Note that the most important input of accent variance adaptor, that is accent intensity, is obtained through the pre-training step, termed as ``accent intensity modeling''. Therefore, in Sec. \ref{subsec:ava}, we will demonstrate the workflow of the accent intensity modeling on the available L2 speech corpus. At last, the run-time inference will be explained.

\subsection{Accent Variance Adaptor}
\label{subsec:ava}
The traditional variance adaptor in \cite{ren2020fastspeech} just adds different variance information such as duration, pitch and energy into the hidden sequence to predict the mel-spectrogram features, that lacks a controlling mechanism. %. This process does not involve any accent effect rendering.
{Next we introduce an accent variance adaptor to provide accent information according to the accent intensity. Specifically, we fuse the L2 speaker identity, accent identity and its accent intensity before the prediction of pitch, energy and duration. 
In other words, the accent variance adaptor learns to project the desired accent and its intensity into the input phoneme embedding $\mathcal{H}_{ph}$. 
This enables all variables, namely the pitch, energy, and duration, to be regulated by the fine-grained accent intensity. Note that the duration predictor is located at the end of the accented variance adaptor instead of the beginning of the variance adaptor in the original FastSpeech2~\cite{ren2020fastspeech}. It was reported that it is more effective to render accent at phoneme-level  than  frame-level~\cite{lancucki2021fastpitch}.}
% since we consider that pitch, energy, and phoneme duration are three major prosodic attributes~\cite{bolinger1958theory} that affect accent rendering. 

As shown in Fig. \ref{fig:overall}(b), the adaptor consists of 1) a L2 speaker encoder, 2) a L2 accent encoder, 3) an intensity encoder, 4) a phoneme pitch predictor, 5) a phoneme energy predictor, and 6) a phoneme duration predictor.

\begin{itemize}
 \vspace{2mm}
\item \textit{L2 Speaker Encoder:} The L2 speaker encoder is a learnable lookup table, that encodes the L2 speaker identity $\mathbf{s}$ into a speaker {code},  $\mathbf{e}_{\mathbf{s}}=E_{spk}(\mathbf{s})$  \cite{gibiansky2017deep,ping2018deep}, to represent a speaker.
% $E_{spk}(\cdot)$, as follows,
% \vspace{-2mm}
% \begin{equation}
%     \mathbf{e}_{\mathbf{s}}=E_{spk}(\mathbf{s})
% \end{equation} 
 \vspace{2mm}
\item  \textit{L2 Accent Encoder:}
Similarly, the L2 accent encoder $E_{acc}(\cdot)$ is another learnable lookup table that transforms an L2 accent category  $\mathbf{a}$ into an accent embedding, %, %\textcolor{red}{(to Rui: is $\mathbf{a}$ an embedding or an utterance? you didn't answer my question. what is  the form of $\mathbf{a}$? is it a vector or a value? or a sentence?)}-----\hl{Dear Prof.Li, a is a value. s also is a value to indicate the speaker id. Speaker encoder and accent encoder convert s and a to embeddings.}----\textcolor{blue}
%{$\mathbf{a}$ is a letter which indicates the accent type. similar to speaker ID. We need to transform it into an meaningful embedding for model training.}
$\mathbf{e}_{\mathbf{a}}=E_{acc}(\mathbf{a})$.
% \vspace{-2mm} 
% \begin{equation}
%     \mathbf{e}_{\mathbf{a}}=E_{acc}(\mathbf{a})
%     \vspace{-1mm} 
% \end{equation}

 \vspace{2mm}
\item \textit{Intensity Encoder:} The intensity encoder $E_{int}(\cdot)$ is a linear layer that transforms a real-valued accent intensity scalar ${i}$ to an embedding vector, $\mathbf{e}_{{i}}=E_{int}({i})$.
% \vspace{-2mm}
% \begin{equation}
%     \mathbf{e}_{{i}}=E_{int}({i})
%     \vspace{-1mm}
% \end{equation}
The real-value scalar of accent intensity is generated by a novel pre-training module, named ``accent intensity modeling'', which will be described in Sec. \ref{subsec:intensity}.

As shown in Fig. \ref{fig:overall}(b),  we concatenate~\cite{im2022emoq} the accent embedding $\mathbf{e}_{\mathbf{a}}$ and its intensity $ \mathbf{e}_{{i}}$, which is then summed with the speaker code $\mathbf{e}_{{s}}$ and the $\mathcal{H}_{ph}$ to form an accented phoneme embedding $\mathcal{H}'_{ph}$. %\textcolor{red}{(1/ this paragraph needs to be re-written. There is no justification of why we sum or concatenate? why sum? why concatenate? this paper is very descriptive, we only describe what we do. but this is not important, what is important is why we design like this. once we explain the design concept, there are multiple ways of implementation.  2/ it is important to justify why we use a variance adaptor first then a predictor if this is our contributions. we can justify under the Section III.A.  )}
%----\textcolor{blue}{Dear Prof.Li, I highlighted the motivation in  subsubsection \ref{subsec:ava}.}
% $L$ denotes the dimension of the $\mathcal{H}_{ph}$.
 \vspace{2mm}
\item  \textit{Variance Predictors:}
With the accented phoneme embeddings, we would like to predict three key prosodic elements that characterize accents, i.e. pitch, energy and duration of phonemes.
The phoneme pitch and energy predictors take $\mathcal{H}'_{ph}$ as input and generate the pitch and energy embedding $\mathbf{p}$, $\mathbf{e}$.
% as shown in Fig. \ref{fig:pr-pred}. 
The two predictors share a similar architecture, that has a 2-layer 1D-convolutional network, a fully connected (FC) layer and an extra 1D-convolutional network. 

To train the predictors, we first extract the frame-level pitch from the voiced speech frame, and set it to 0 for unvoiced frames.
% {Unlike in FastSpeech2~\cite{ren2020fastspeech} where frame-wise pitch is quantized to 256 possible values in log-scale, we use linear scale to represent the pitch scalar directly.}
We compute the L2-norm of the amplitude of each short-time Fourier transform (STFT) frame as the energy. We obtain frame-level pitch and energy averages for each phoneme using the ground-truth duration $\mathcal{D}$. 
The phoneme pitch and energy scalars are normalized to zero mean and unit variance over all training data, to serve as the supervision signals. 
% As shown in Fig. \ref{fig:pr-pred}, 
We define mean square error loss $\mathcal{L}_{p\_pitch}$ and $\mathcal{L}_{p\_energy}$ behind the FC layers as the objective functions for the training of the pitch and energy predictors.

% The phoneme duration predictor share a similar architecture with that of FastSpeech2 .
Unlike in FastSpeech2 \cite{ren2020fastspeech} where the phoneme duration predictor takes the encoder output directly as input, our phoneme duration predictor takes the accented phoneme embedding $\mathcal{H}^{'}_{ph}$ as input and is expected to output more accurate phoneme duration $\mathcal{D}$. At last, we sum the accented phoneme, pitch and energy embeddings to form an accented phoneme embedding $\mathcal{H}^{''}_{ph} = \mathcal{H}^{'}_{ph} + \mathbf{p} +\mathbf{e}$. 
A length regulator (LR) is used to transform the $\mathcal{H}^{''}_{ph}$  to frame-level embeddings  $\mathcal{H}_{fm}$ based on the predicted phoneme duration $\mathcal{D}$. Finally, the mel-spectrum decoder converts the $\mathcal{H}_{fm}$, along with a triangle positional embedding $P\!E$ \cite{vaswani2017attention}, into a mel-spectrum sequence. % in parallel.

\end{itemize}

\subsection{Consistency Constraint for Accent Intensity
}

%Consistency constraint for accent intensity aims to force the estimated accent intensity, which is predicted from mel-spectrum, close to the intensity score obtained from accent intensity learning, as illustrated in Sec. \ref{subsec:intensity}.

The consistency constraint module ensures the synthesized L2-accented speech manifests the expected accent intensity at a fine level, as shown in Fig. \ref{fig:overall}(a). The consistency constraint is applied via a consistency constraint loss $\mathcal{L}_{cc}$ between the intended accent intensity $i$ and the accent intensity $\hat{i}$ measured by the accent intensity predictor, i.e., $\mathcal{L}_{cc} = {\rm MSE}(i, \hat{i})$.

The accent intensity predictor consists of a bi-directional recurrent neural network layer with Gated Recurrent Unit (GRU), that is followed by an FC layer. It is trained in a supervised manner together with the CAI-TTS model subject to the total loss, $\mathcal{L}_{final} = \mathcal{L}_{mel} + \mathcal{L}_{dur} + \mathcal{L}_{p\_pitch} + \mathcal{L}_{p\_energy} + \mathcal{L}_{cc}$, where $\mathcal{L}_{mel}$ and $\mathcal{L}_{dur}$ are the MSE loss for mel-spectrum and duration loss as in FastSpeech2 \cite{ren2020fastspeech}.

% to absorb the forward and backward context information, following by an FC layer.
%It takes the predicted mel-spectrum sequence, output by mel-spectrum decoder, as input and predicts the accent intensity $\hat{i}$.To force the predicted accent intensity $\hat{i}$ and the ground truth accent intensity $i$ close to each other, we define a consistency constraint loss $\mathcal{L}_{cc}$ to compare the distance between them: Under the supervision of consistency constraint loss, the CAI-TTS model learns to project the accent intensity accurately.

%During tra the consistency constraint loss with the conventional loss function during training. The final loss function is formulated as follows:
% \vspace{-3mm}

% \begin{equation}
%   \mathcal{L}_{final} = \mathcal{L}_{mel} + \mathcal{L}_{dur} + \mathcal{L}_{p\_pitch} + \mathcal{L}_{p\_energy} + \mathcal{L}_{cc}
% \end{equation}

 %; $\mathcal{L}_{p\_pitch}$ and $\mathcal{L}_{p\_energy}$ mean the MSE loss for the phoneme pitch and energy scalars, respectively, in this paper.

% \subsubsection{Run-time Inference}
% During inference, the CAI-TTS system takes a phoneme sequence as the input text. We specify the intended L2 speaker ID, the accent ID and its intensity scalar, e.g, (`BWC' `Mandarin', 0.8) as the controlling factors to generate the accented speech waveform.

%\textcolor{black}{(to Rui Liu  please confirm. I see Fig.1a has phoneme sequence as input? please explain in Section 3.1)} 

\begin{figure}[t]
    \centering
    % \vspace{-4mm}
    \centerline{\includegraphics[width=0.6\linewidth]{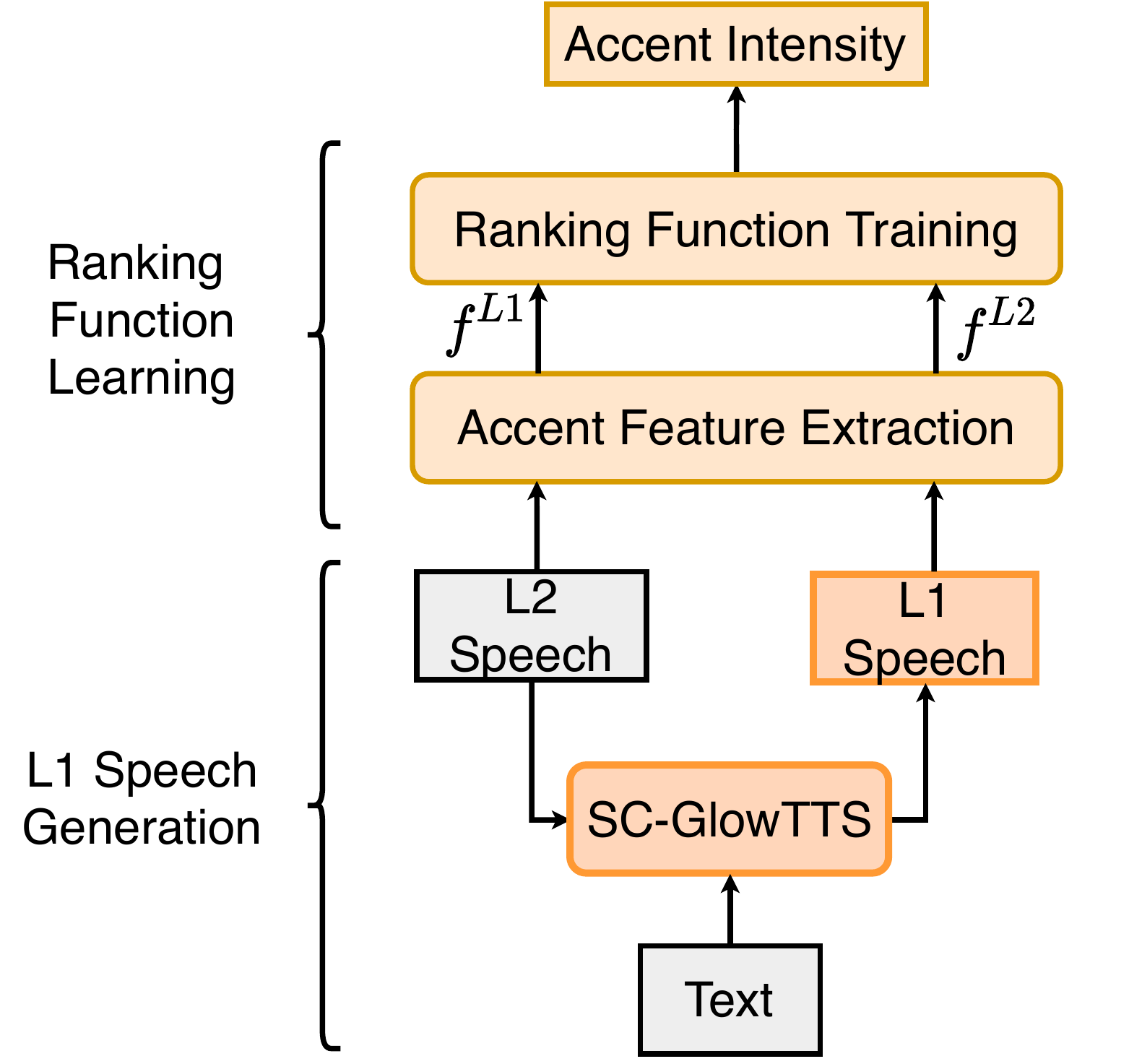}}
    % \vspace{-3mm}
    \caption{The workflow of the accent intensity modeling. The accented TTS corpus includes $<$Text, L2 Speech$>$ pairs.}
    \label{fig:intenlearn}
    % \vspace{-5mm}
\end{figure}

\subsection{Accent Intensity Modeling}
\label{subsec:intensity}
 
The key to accent control at a fine level is to find a descriptor to quantify the accent intensity. To this end, we propose a novel pre-training strategy, named ``accent intensity modeling'' scheme, that seeks to quantify the accent intensity for each sample of L2-accented speech before CAI-TTS training, as illustrated in Fig. \ref{fig:intenlearn}. The accent intensity modeling scheme consists of \textit{L1 speech generation} and \textit{ranking function learning}. The ranking function learning includes \textit{accent feature extraction} and \textit{ranking function training}.

Note that L1 speech generation aims to develop a $<$L1, L2$>$ paired speech dataset, that is used for the ranking function learning. Then the ranking function learning will learn the difference between L1 and L2 speech pair to quantify the accent intensity for L2 speech. We will introduce the workflow in order.

%In ranking function learning, the paired speech samples will go to an accent feature extractor to extract the pairwise acoustic features with our designed accent feature set.
%Next, we try to train a ranking function from the paired acoustic features.
%At last, we can use the trained ranking function to estimate a real-valued score (range from 0 to 1), for each accent speech sample of the accent TTS corpus, as the accent intensity.

%We will introduce the L1 speech generation and ranking function learning processes in the following subsections.

\subsubsection{L1 Speech Generation}
 
\begin{figure}[]
    \centering
    % \vspace{-4mm}
   \centerline{
    \includegraphics[width=0.7\linewidth]{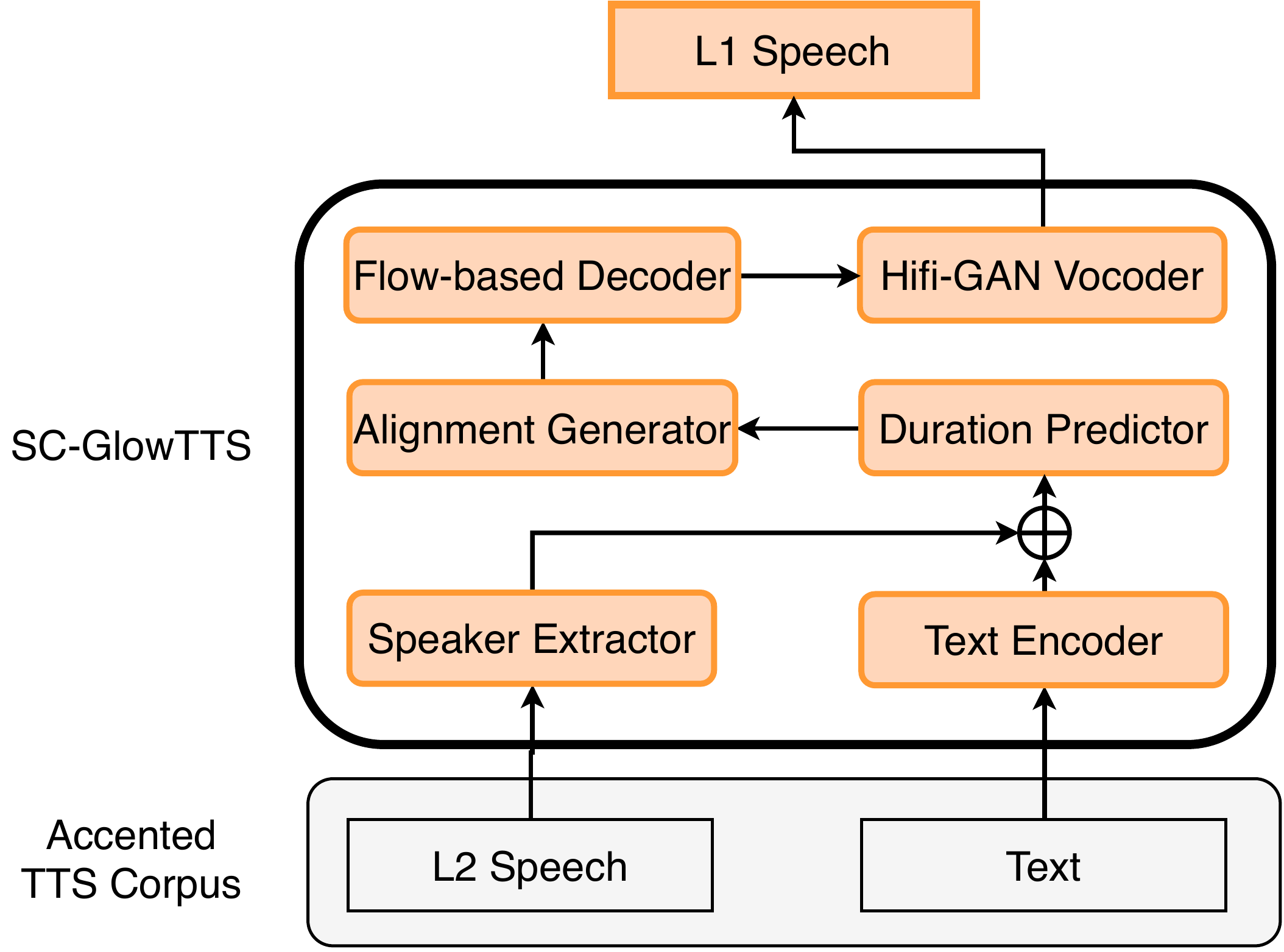}
}
    \caption{L1 speech generation using SC-GlowTTS.}
    \label{fig:scglowTTS}
    % \vspace{-5mm}
\end{figure}

%To model the accent intensity through ranking function learning, 
We first create a  $<$L1, L2$>$ parallel speech database of the same speech content, where an L1 speech sample is synthesized for each L2 utterance of the same speaker by SC-GlowTTS \cite{casanova2021sc}, a state-of-the-art zero-shot TTS model as in Fig. \ref{fig:scglowTTS}.  The high-quality L1 speech samples serve as the reference for the ranking function learning. % We follow ~\cite{casanova2021sc} for SC-GlowTTS implementation, which consists of an external speaker extractor, and adopts a flow-based non-autoregressive model for mel-spectrum generation. Finally, a HiFi-GAN \cite{kong2020hifi} vocoder is used to convert the mel-spectrum to speech waveform. %. We synthesize a L1 speech sample for the same speaker identity, and speech content of each L2-accented speech sample. 

%are extracted from the accented TTS corpus, and  for the accented TTS corpus, the most important is how to determine the difference between the accented L2 and its corresponding L1 speech. 
% ‘Unfortunately, the accented TTS corpus just provides accented L2 speech samples, without L1 reference. 
%The first step is how to preserve the L2 speaker identity, at the same time, generate high-quality L1 speech?
    
%Fortunately, zero-shot TTS \cite{arik2018neural}, which can synthesize high-quality voice for new speakers (unseen during training), opens the door for us to solve this problem. Recently, a novel zero-shot TTS model, named speaker conditional GlowTTS (SC-GlowTTS)~\cite{casanova2021sc}, was proposed. SC-GlowTTS uses an external speaker encoder based on Angular Prototypical loss~\cite{chung2020defence}, to learn speaker embedding vectors, and adapt the HiFi-GAN \cite{kong2020hifi} vocoder to convert the output spectrum to the waveform. Note that SC-GlowTTS reaches state-of-the-art results for similarity with new speakers, as well as high speech quality \cite{casanova2021sc}. 

% The work flow is illustrated in Fig. \ref{fig:scglowTTS} of Appendix A.

{As in \cite{casanova2021sc}, the SC-GlowTTS model employs an external speaker extractor, that encodes an utterance into a  speaker embedding. We adopt the flow-based non-autoregressive model to predict the mel-spectrum sequence. 
Finally, a HiFi-GAN \cite{kong2020hifi} vocoder converts the output mel-spectrum sequence to speech waveform.  }

{    
In practice, for a text-speech pair in the accented TTS corpus, 
% the text transcript is first converted to a phoneme sequence $\mathcal{P}$ by the grapheme to phoneme (G2P) Conversion toolkit \footnote{https://github.com/Kyubyong/g2p}. In this way, 
we first obtain a phoneme-speech pair $\{\mathcal{P}, Y\}$ where $\mathcal{P}= \{\mathcal{P}_{1}, \mathcal{P}_{2}, ..., \mathcal{P}_{T}\}$ represents a phoneme sequence of $T$ tokens, while $Y = \{y_{1}, y_{2}, ..., y_{N}\}$ represents the L2 speech of $N$ samples.}
The text encoder takes the phoneme sequence $\mathcal{P}$ as input, while the speaker extractor encodes a L2 speech utterance into a {speaker embedding.} In this way, the system generates L1 speech in L2 speaker's voice. %\textcolor{red}{Finally, we obtain the $<$L1, L2$>$ paired speech samples, which share the  same content and speaker identify but in different accents.}

% In detail, given the accented TTS corpus $D = \{X, Y\}$, which $X = \{x_{1}, x_{2}, ..., x_{T}\}$ ($T$ denotes the total number of samples in $D$) means the text transcription and $Y = \{y_{1}, y_{2}, ..., y_{T}\}$  means the accent audio samples, the text transcription $x_{i}$ ($i \in [1,T]$) is first converted to a phoneme sequence $P_{i} = \{p_{i1}, p_{i2}, ..., p_{iN}\}$  using the grapheme to phoneme (G2P) Conversion toolkit \footnote{https://github.com/Kyubyong/g2p}.

% The SC-GlowTTS takes the phoneme sequence $P_{i}$ and the L2 speech $y_{i}$ as input and outputs the synthesized L1 speech $\hat{y}_{i}$ with L2 speaker.

%In such a way, we obtain the $<$L1, L2$>$ pairwise speech samples, that is $<\!\!Y, \hat{Y}\!\!>$. 
% \textcolor{black}{why you need to introduce $<\!\!Y, \hat{Y}\!\!>$, please just use $<$L1, L2$>$}
% in which $\hat{Y} = \{\hat{y}_{1}, \hat{y}_{2}, ... \hat{y}_{T}\}$.
%The $<$L1, L2$>$ paired speech samples will go to ranking function learning module, including accent feature extraction and ranking function training, to learn a ranking function to estimate the accent intensity for accented L2 samples.

  \vspace{2mm}
\subsubsection{Ranking Function Learning}
 
In ranking function learning, we first extract accent acoustic features $<\!\!f^{L1}, f^{L2}\!\!>$ from a pair of speech samples $<$L1, L2$>$, then train the ranking function, following the attribute ranking idea as described in Sec. \ref{subsec:rank}.

\vspace{2mm}
\textit{2.1) Accent Feature Extraction:}  
Studies show that pitch and energy~\cite{bolinger1958theory} related features are the descriptors of accents. 
% Specifically, {low-level descriptor contours (e.g. Mel frequency cepstral coefficients, signal frame energy, zero-crossing rate, fundamental frequency) and their statistical functionals (e.g. maximum, minimum, range, position, arithmetic mean, standard deviation, skewness, kurtosis, linear regression slope etc.)} 
We use opensmile \cite{eyben2010opensmile} to extract 36-dimensional features  $f^{L1}$ and $f^{L2}$ for each utterance, each of which contains both pitch and energy features.
% = \{\hat{f}_{1}, \hat{f}_{2}, ..., \hat{f}_{T}\}$
% for L2 and synthesized L1 speech respectively, resulting in pairwise feature sequence $<\!\!f, \hat{f}\!\!>$ for ranking function training.
 
 \vspace{2mm}
\textit{2.2) Ranking Function Training:}  
Motivated by \cite{parikh2011relative}, we would like to learn a ranking function $R(\cdot)$ from the paired feature sequence $<\!\!f^{L1}, f^{L2}\!\!>$ by using relative attribute ranking for the accent attribute,
% We use $\mathcal{F} = f^{L1}  \cup  f^{L2}$ to denote the union of $f^{L2}$ and $f^{L2}$, and $\mathcal{F}_{m}$ to represent the feature sequence of $m$th ($m \in [1,2T]$, $T$ denotes the total number of samples in accented TTS corpus) sample of $\mathcal{F}$.Our goal is to learn a ranking function $R(\cdot)$.
 
\begin{equation}
    i = R(f^{L2}_{m}) = \textbf{w} f^{L2}_{m}
\end{equation}
where we weight $f^{L2}_{m}$ with a learnable weighting vector $\textbf{w}$ and return a weighted sum $i$ indicating the accent intensity of sample $m$. %$\textbf{w}$ means a learnable weighting  vector. %-----Dear Prof.Li, I checked the code and confirm that $w$ is a vector. The shape of $f^{L2}_{m}$ is [1,36], while the $w$ is [36,1]. Sorry about that.}.
%\textcolor{black}{1/ if $w$ is a matrix, it should be $W$. 2/ But  $R(f^{L2}_{m}) = w f^{L2}_{m}$ is wrong, because left handside is a scalar value, and right hand size is a vector - a matrix multiplies a vector = vector. -----------------------Dear Prof.Li, I add a Transpose symbol for $w$.The shapes of $w^{\top}$ and $f^{L2}_{m}$ are transposed, so their product is a scalar. if w is a matrix, f is a vector, the result is always a vector! do you mean w is a vector? you wrote w is a learnable weighting matrix???-----------Sorry about this, original paper just use $w^{\top}$. maybe it's a vector.it cannot be `maybe', please check the codes.--ok,I will check now.}

To achieve this, we first build two sets, that are $O$ and $S$, which contains ordered and similar paired samples respectively,
according to the following relative constraints:

\begin{equation}
\left\{\begin{matrix}
 \forall (m,h) \in O: \textbf{w}   \mathcal{F}_{m}> \textbf{w} \mathcal{F}_{h}   \\ 
 \forall (m,h) \in S: \textbf{w}   \mathcal{F}_{m}= \textbf{w} \mathcal{F}_{h}
\end{matrix}\right.
\end{equation}
where $\mathcal{F}$ represents a sample from the union of $f^{L2}$ and $f^{L2}$: $\mathcal{F} \in \{f^{L1}  \cup  f^{L2}\}$.
 
Specifically, we pick up one sample $\mathcal{F}_{m}$ from $f^{L1}$ and another sample $\mathcal{F}_{h}$ from $f^{L2}$ to build the ordered set $O$.
We expect that the accent intensity of the L2-accented sample $\mathcal{F}_{h}$ is higher than that of the L1 sample $\mathcal{F}_{m}$.
For the similar set $S$, we pick up two samples $\mathcal{F}_{m}$ and $\mathcal{F}_{h}$ from $f^{L1}$ (or $f^{L2}$). We assume that two samples from the same domain (L1 or L2) have similar accent intensities.

% The learning process of such a ranking function can be formulated as a max-margin optimization problem \cite{chechik2008max} and solved by the Newton's method \cite{chapelle2007training}. 

Next, we can learn a support vector machines (SVM) \cite{joachims2002optimizing} to estimate $\textbf{w}$ by solving the following optimization problem:

\begin{small}
\begin{equation}
\setlength{\abovedisplayskip}{-1pt}
\setlength{\belowdisplayskip}{-1pt}
% \vspace{-5mm}
    {\rm minimize}  \left(\frac{1}{2} ||\textbf{w}||^{2}_{2} + C \left( \sum {\xi^{2}_{ij}} + \sum {\eta ^{2}_{ij} } \right) \right)
\end{equation}
% \vspace{-2mm}
\end{small}

All parameters need to satisfy the following constraints:

\begin{equation}
\setlength{\abovedisplayskip}{-1pt}
\setlength{\belowdisplayskip}{5pt}
     \left\{\begin{matrix}
 w(\mathcal{F}_{m}-\mathcal{F}_{h}) \geqslant  1-\xi_{mh} ; \forall (m,h)\in O\\
|w(\mathcal{F}_{m}-\mathcal{F}_{h})| \leqslant   \eta_{mh} ; \forall (m,h)\in S\\
 \xi_{mh}\geqslant 0;\eta_{mh} \geqslant 0
\end{matrix}\right.
\end{equation}
where $C$ is to control the trade-off between the margin and the size of the slack variables $\xi_{mh}$ and $\eta _{mh}$.
The use of $\xi_{mh}$ and $\eta _{mh}$ is to relax the constraints on classifying all of the ordered and similar samples in $O$ and $S$, respectively.
This primal problem can be solved by the Newton's method \cite{chapelle2007training}.
Note that the intensity scalar $i$ is normalized to (0,1) with 1 as the highest intensity.

\subsection{Run-time Inference}

At run-time, the CAI-TTS architecture takes Input text, L2 speaker ID, L2 accent ID and the custom accent intensity as input to generate the output L2 speech with the help of universal HiFi-GAN vocoder \cite{kong2020hifi} in this paper.

Note that the predicted pitch contour and energy also can be modified during inference to control certain perceived qualities of the generated speech like FastPitch \cite{lancucki2021fastpitch} etc. 
However, synthesized speech with a large pitch (or energy) shift scale suffers from audio quality degradation, and speaker characteristics deformation \cite{bak2021fastpitchformant}.
In our work, we just try to modify the L2 speaker ID, accent ID and accent intensity to achieve accented and intensity controllable TTS synthesis.

\section{Experiments}
\label{sec:exp}
 
We evaluate CAI-TTS through accented text-to-speech synthesis experiments \footnote{Audio samples: https://speechdemo.github.io/caitts}.

\subsection{Datasets}
We train CAI-TTS on the publicly available L2-ARCTIC corpus \cite{zhao2018l2}, {which includes about 26 hours recordings of accented English from 24 non-native speakers, whose are native in Hindi, Korean, Mandarin, Spanish, Arabic and Vietnamese. Two male and two female speakers contributed in each language. }

In L2-ARCTIC, scripts and their phoneme-level duration annotations are provided. 
The scripts for each speaker consists of about 1,130 utterances, resulting in about 27,120 utterances in total. 
Its phonetic transcription follows the ARPAbet phoneme set \footnote{http://www.speech.cs.cmu.edu/cgi-bin/cmudict}.
The speech data are sampled at 44.10kHz and coded in 16 bits. For each accented speaker, we partition the speech data into training, validation, and test set at a ratio of 8:1:1.

\subsection{Comparative Study}
 
This work is one of the first attempts to control multiple accents and their intensity at a fine level. As there is no reference system in the literature, we choose some state-of-the-art TTS systems as the benchmark.
\textbf{1) GT (\textit{oracle})}: This is the natural L2 speech by the speakers;
\textbf{2) GT mel + HiFi-GAN (\textit{oracle})}: This is the synthesized speech with a HiFi-GAN vocoder using GT mel-spectrum;
\textbf{3) Tacotron2}: This is a multi-speaker Tacotron2 \cite{shen2018natural} model which learns a look-up-table to map embeddings for different speaker identity;
\textbf{4) Transformer-TTS}: This is a multi-speaker Transformer-TTS \cite{chen2020multispeech} model which learns a look-up-table to map embeddings for different speaker identity;
\textbf{5) FastSpeech2}: This is a multi-speaker FastSpeech2 \cite{ren2020fastspeech} model which learns a look-up-table to map embeddings for different speaker identity;
\textbf{6) CAI-TTS (\textit{proposed})}: This is our proposed model. % which includes a novel accent variance adaptor and a consistency constraint module to achieve multi-accent TTS synthesis with intensity controllable. 
% \begin{itemize}
%     \item \textbf{GT (\textit{oracle})}: This is Ground-Truth accented L2 speech;
%     \item \textbf{GT mel + Vocoder (\textit{oracle})}: This is the L2 speech synthesized by our Hifi-GAN vocoder using Ground-Truth mel-spectrum;
%     \item \textbf{Tacotron2}: This is a multi-speaker Tacotron2 \cite{shen2018natural} model which learns a look-up-table to map embeddings for different speaker identity;
%     \item \textbf{Transformer-TTS}: This is a multi-speaker Transformer-TTS \cite{chen2020multispeech} model which learns a look-up-table to map embeddings for different speaker identity;
%     \item \textbf{FastSpeech2}: This is a multi-speaker FastSpeech2 \cite{ren2020fastspeech} model which learns a look-up-table to map embeddings for different speaker identity;
%     \item  \textbf{CAI-TTS (\textit{proposed})}: This is our proposed model which includes a novel accent variance adaptor and a consistency constraint module to achieve multi-accent TTS synthesis with intensity controllable. 
% \end{itemize}
We also develop 3 variants of CAI-TTS model for an ablation study.
\textbf{7) CAI-TTS w/o accent intensity}: We replace the accent variance adaptor with the variance adaptor of FastSpeech2 \cite{ren2020fastspeech};
\textbf{8) CAI-TTS w/o phoneme pitch \& energy}: We replace the phoneme level pitch and energy predictors with frame-level pitch and energy predictors as in  \cite{ren2020fastspeech};
\textbf{9) CAI-TTS w/o consistency constraint}: We remove the consistency constraint module from the CAI-TTS model.

\subsection{Experimental Setup}
 
The text encoder and mel-spectrum decoder use 6 Feed-Forward Transformer (FFT) blocks. %, which different form the 4 blocks in FastSpeech2 \cite{ren2020fastspeech}.
Note that the text encoder takes the 256 dimensional phoneme embedding as input. 
% The phoneme sequence is generated by  the grapheme to phoneme (G2P) conversion toolkit \footnote{https://github.com/Kyubyong/g2p}.
The mel-spectrum decoder generates 80-channel mel-spectrum, which is extracted with 12.5ms frame shift and 50ms frame length, as output.
We downsample all speech files to 22.05kHz and trimmed leading and trailing silence.
% using Librosa \footnote{https://github.com/librosa}.
% The 80-channel mel-spectrum is extracted with 12.5ms frame shift and 50ms frame length.

The L2 speaker and accent encoders employ two lookup tables with 14$\times$256 and 6$\times$128 respectively.  
The intensity encoder consists of a linear layer, which encodes the accent intensity scalar into a 128 dimensional intensity embedding.

Following \cite{lancucki2021fastpitch}, we represent the pitch and energy scalar in linear scale.
The pitch and energy predictors employ a Conv1D with kernel size 3 and 384/256 channels, Conv1D with 256/256 channels, and an FC layer to project a 256-channel vector into a single pitch/energy scalar, and the last Conv1D layer with kernel size 9 to upsamples pitch/energy scalar to pitch/energy embeddings.
% \textcolor{black}{why real-value here?}
% Unlike in FastSpeech2~\cite{ren2020fastspeech} where a 256-dimensional embedding is used to represent pitch and energy, \textcolor{black}{i don't understand the next sentence! what is the pitch information? do you mean the scalar supervision value? }the last Conv1D layer is used to add pitch and energy information directly to the accented phoneme embedding.
We use Dropout rate of 0.5 in every Dropout layer.
% \textcolor{black}{in the experiment, we should only talk about parameter setting, if you are discussing the architecture, the text should be moved to Section III. please DO NOT REPEAT the same sentence in the paper 2 times!}
% \textcolor{blue}{The phoneme pitch and energy scalars are normalized by mean and std values over all training set, to serve as the reference target for phoneme pitch and energy predictors.}
% The duration predictor share a same architecture with FastSpeech2 \cite{ren2020fastspeech} while locate at the last position in the variance adaptor.
The accent intensity predictor in consistency constraint module consists of a GRU layer with hidden size 128  and an FC layer to output the intensity scalar.

We use the Adam optimizer \cite{kingma2014adam} with $\beta_1$ = 0.9, $\beta_2$ = 0.98 and follow the same learning rate schedule in \cite{vaswani2017attention}.
All models are trained with 900k steps to ensure complete convergence.
The codes are written in Python 3.6 using the Pytorch library 1.7.0. The GPU type is NVIDIA Quadro RTX 6000 with 24GB GPU memory. 
For a fair comparison, we adopt the same pretrained universal HiFI-GAN \cite{kong2020hifi} vocoder for all systems.
As the universal HiFI-GAN vocoder is trained on a collection of multiple dataset, that includes LJSpeech \cite{ljspeech17}, VCTK \cite{yamagishi2019cstr} and LibriTTS \cite{zen2019libritts}, it is known to produce high-quality voice for unseen speakers~\cite{lorenzo2019towards,jiao2021universal}.

\begin{table*}[!t]
\centering
\small
\renewcommand\arraystretch{1.2}
\caption {The comparison of the audio quality for different systems in terms of MCD in objective experiments, MOS and BWS in subjective experiments.
}
 
\begingroup
\begin{tabular}{p{6.8cm}|p{1.5cm}<{\centering}|p{2.5cm}<{\centering}|p{1.5cm}<{\centering}p{1.5cm}<{\centering}}
% \begin{tabular}{l|c|c}
\toprule
\multirow{3}{*}{$\qquad \qquad \qquad \qquad \qquad$System} & \multicolumn{4}{c}{Audio Quality}  \\ \cline{2-5}
 & \multirow{2}{*}{\textit{MCD} (dB)} & \multirow{2}{*}{\textit{MOS}} & \multicolumn{2}{c}{BWS Evaluation}       \\
\cline{4-5}
& & & \textit{Best} (\%) & \textit{Worst} (\%)\\ \hline
GT   & NA  & 4.62 $\pm$ 0.02 & NA & NA \\
 GT (Mel + HiFi-GAN)  & NA  & 4.59 $\pm$ 0.03 & NA & NA
\\ \hline 
  Tacotron2 \cite{shen2018natural} (Mel + HiFi-GAN)  & 6.57  & 4.33 $\pm$ 0.01 & 6 & 34
\\ 
Transformer TTS \cite{li2019transformerTTS} (Mel + HiFi-GAN) & 6.43 & 4.35 $\pm$ 0.02 & 8 & 22
\\  
 FastSpeech2 \cite{ren2019fastspeech} (Mel + HiFi-GAN)   & 6.31  & 4.38 $\pm$ 0.03 & 5 & 18
\\ \hline 
\textbf{CAI-TTS} \textit{\textbf{(proposed)}} (Mel + HiFi-GAN)  & \textbf{6.08} & \textbf{4.57 $\pm$ 0.01} & \textbf{45} & \textbf{4
}\\ \hline 
 \quad  w/o phoneme pitch \& energy  & 6.11  & 4.52 $\pm$ 0.02 & 9 & 8
\\  
 \quad  w/o accent intensity  & 6.14  & 4.48 $\pm$ 0.03 & 13 & 6
\\  
\quad  w/o consistency constraint  & 6.16  & 4.43 $\pm$ 0.01 & 14 & 8\\ 
% \hline
% \quad  w/o accent variance adaptor \&  consistency constraint  & xx $\pm$ xx  & xx $\pm$ xx & xx & xx\\
\bottomrule
\end{tabular}
 
\endgroup
\label{tab:tab1}
\end{table*}

\begin{table*}[!h]
\centering
\small
\renewcommand\arraystretch{1.2}
\caption {The comparison of the accent variance information for different systems, including standard deviation ($\sigma$), skewness ($\gamma$), kurtosis ($\mathcal{K}$) and average dynamic time warping (DTW) distances ($\varrho$) for pitch, mean absolute error (\textit{MAE}) for energy and the average of absolute boundary differences ($\Delta$) for duration.
}
 
\begingroup
\begin{tabular}{p{6.8cm}|p{1.05cm}<{\centering}p{1.05cm}<{\centering}p{1.05cm}<{\centering}p{1.05cm}<{\centering}|p{1.05cm}<{\centering}|p{1.05cm}<{\centering}}
% \begin{tabular}{l|cccc|c|c}
\toprule
\multirow{3}{*}{$\qquad \qquad \qquad \qquad \qquad$System} & \multicolumn{6}{c}{Accent Variance}  \\ \cline{2-7}
   & \multicolumn{4}{c|}{Pitch} & Energy & Duration  \\ \cline{2-7}
  & $\sigma$ &  $\gamma$ &  $\mathcal{K}$ & $\varrho$  & \textit{MAE} &  $\Delta$ (ms)       \\
\hline 
GT & 49.8  & 0.627  & 0.854  & NA & NA & NA
\\
 GT (Mel + HiFi-GAN)  & 49.5  & 0.629  & 0.852 & 15.27  & 0.192 & NA
\\ \hline 
  Tacotron2 \cite{shen2018natural} (Mel + HiFi-GAN)   & 32.5 & 0.906  & 1.212  & 19.80 & 0.297 & 22.12
\\ 
Transformer TTS \cite{li2019transformerTTS} (Mel + HiFi-GAN)  & 35.8  &  0.887 & 1.025 & 19.32  & 0.288 & 21.98
\\  
 FastSpeech2 \cite{ren2019fastspeech} (Mel + HiFi-GAN)  & 43.2 & 0.752 & 0.973 &  18.01 & 0.274 & 20.08
\\ \hline 
\textbf{CAI-TTS} \textit{\textbf{(proposed)}} (Mel + HiFi-GAN)  & \textbf{48.9} & \textbf{0.613}  & \textbf{0.882} & \textbf{16.78} & \textbf{0.235} & \textbf{18.54}
\\ \hline 
 \quad  w/o phoneme pitch \& energy  & 48.3  & 0.610 & 0.896 & 17.06 & 0.240 & 18.72
\\  
 \quad  w/o accent intensity  & 47.9 & 0.611 & 0.890 & 16.95 & 0.239 & 18.70
\\ 
\quad  w/o consistency constraint  & 47.8  & 0.594 & 0.889& 17.13& 0.245 & 18.85\\ 
% \hline
% \quad  w/o accent variance adaptor \& consistency constraint  & xx $\pm$ xx  & xx $\pm$ xx & xx & xx\\
\bottomrule
\end{tabular}
 
\endgroup
\label{tab:tab2}
\end{table*}

%Before investigating the ability of our CAI-TTS model, we first evaluate the performance of the zero-shot TTS for L1 speech generation. The details of the subjective evaluations for the L1 speech generation are added in Appendix C.

\subsection{Audio Quality of Generated L1 Speech}

% \textcolor{black}{please keep just 1 paragraph for this subsection. we need to show that we have evaluated the quality of L1 speech generation.}
The accent intensity modeling cannot be done well without L1 speech with high audio quality. Therefore, we first conduct a Mean Opinion Score (MOS) and a Speaker Similarity MOS (SMOS) evaluation to validate the overall performance of the L1 speech generated by zero-shot TTS in terms of audio quality and speaker similarity.

% The details of both experiments are given in Appendix B.
% In order to evaluate the effectiveness of zero-shot TTS for L1 speech generation, we conduct subjective experiments to test the overall performance of the synthesized L1 speech.
\textcolor{black}{Specifically, we conduct listening experiments with mean opinion score (MOS) for audio quality and similarity MOS (SMOS) \cite{lorenzo2018voice} for speaker similarity.
Both metrics are rated in 1-to-5 scale and reported with 95\% confidence intervals (CI).}

\textcolor{black}{
We first conduct MOS listening experiments to compare the overall audio quality between L2 speech and synthesized L1 speech.
100 $<$L1, L2$>$ pairwise sentences were chosen as the test samples randomly. We invite 20 subjects to rate the audio quality for each sample. Each listener listened to 200 speech samples.
The MOS values are calculated by taking the arithmetic average of all scores provided by the subjects. 
To this end, the MOS score of L2 speech reaches 4.57 $\pm$ 0.03, and the MOS score of synthesized L1 speech achieves 4.32 $\pm$ 0.02. Note that  both MOS scores are higher than 4.0, indicating that the quality of synthesized L1 speech is good enough to serve as the reference speech for accent intensity modeling.}

\textcolor{black}{
We also evaluate the speaker similarity between L1 and L2 speech. 
We follow the MOS experiment settings, except that we randomly choose 100 samples as reference samples which share same speaker identity with the test samples.
20 listeners are invited to listen to the reference samples and test samples, and rate the speaker similarity for each test sample. 
After that, we calculate the average SMOS according to all scores.  
Note that the SMOS scores of L1 and L2 speeches got 4.25 $\pm$ 0.03 and 4.49 $\pm$ 0.01 respectively, which are very similar to the trending of MOS scores. We believe that the synthesized L1 speech generated by zero-shot TTS performs enough speaker similarity.}

\textcolor{black}{
In a nutshell, the synthesized L1 speech achieves comparable scores with L2 speech on both experiments in terms of audio quality and speaker similarity, which shows the effectiveness of the zero-shot TTS model for L1 speech generation.
In other words, we observe that the quality of the synthesized L1 speech is on par with the L2-accented speech in both experiments. Therefore, we confirm that the $<$L1, L2$>$ paired speech samples are adequate for accent intensive learning study.}

\subsection{Audio Quality of CAI-TTS}
In this section, we conduct objective and subjection evaluations to validate the overall audio quality of our CAI-TTS model in terms of accented expression.

In the objective evaluation, we first employ Mel Cepstral Distortion (MCD) to measure the spectral distance between the synthesized and reference Mel-spectrum features.
Since the sequences are not aligned, we perform Dynamic Time Warping (DTW) algorithm \cite{muller2007dynamic} to align the sequences prior to comparison.
Lower MCD value indicates smaller distortion, thus better quality.
We randomly select 100 utterances from the test set as test samples and report the results in the second column of Table \ref{tab:tab1}. We observe that the CAI-TTS outperforms all other systems with the lowest MCD of 6.08 dB. In the ablation study, we observe that the three ablation systems see a performance degradation to some extent, which reaffirms the effectiveness of the individual prosody-controlling modules.

In the subjective experiment, we first evaluate all systems in terms of MOS.
We choose 100 utterance as the test samples. 20 listeners were participated and each listener listens to 100 speech samples.
As shown in the third column of Table \ref{tab:tab1}, it is observed that our CAI-TTS achieves a MOS of 4.57 $\pm$ 0.01, that is significantly higher than others and very close to those of the \textit{GT (Mel+HiFi-GAN)} and \textit{GT}.

We conduct the second listening experiment through Best Worst Scaling (BWS) \cite{lee2008best}, which is an effective method to provide a ranking of a long list of listening samples.
BWS evaluation share same test samples with previous MOS evaluation.
For each utterance, seven speech samples produced by these seven  TTS systems (expect \textit{GT(Mel+HiFi-GAN)} and \textit{GT})) form a group. 20 listeners were participated and each listener picks the best and worst samples in terms of naturalness for each group.
We report the results in the last two columns of Table \ref{tab:tab1}. We observe that CAI-TTS is selected for 45\% of time as the best model and 4\% as the worst model. All baselines get lower best percentage and higher worst percentage. 
The results suggest that the listeners have a clear preference towards our proposed CAI-TTS system in terms of accent similarity.

\subsection{Accent Variance Learning}

To understand how the accent variance adaptor performs, we randomly select 100 utterances from the test set as the test samples.

% \vspace{-2mm}
% \subsubsection{Pitch}
% \vspace{-1.4mm}

\noindent\textbf{Pitch}: We follow \cite{ren2020fastspeech} and compute the moments (including standard deviation ($\sigma$), skewness ($\gamma$) and kurtosis ($\mathcal{K}$)), and average DTW distance ($\varrho$) of the pitch distribution between the synthesized L2-accented speech and the ground truth reference.
The results are summarized in the second to fifth columns of Table \ref{tab:tab2}. It can be seen that the CAI-TTS system is reported with the moments ($\sigma$, $\gamma$ and $\mathcal{K}$) that are closer to those of the natural speech (GT) than Tacotron2, Transformer TTS and FastSpeech2. In terms of the average DTW distance $\varrho$ to the ground truth pitch, the CAI-TTS system outperforms all baselines, with the lowest value of 16.78.

% \vspace{-2mm}
% \subsubsection{Energy}  
% \vspace{-2mm}
\noindent\textbf{Energy}: We also compute the accuracy by calculating the mean absolute error (MAE) between the frame-level energy extracted from the synthesized and the ground-truth L2-accented speech. The DTW algorithm is applied to align the paired sequences.
As shown in the sixth column of Table \ref{tab:tab2}, the CAI-TTS system presents the lowest MAE among all benchmarking baselines. %there is no doubt that \textit{GT(mel+HiFiGAN)} achieves the lowest value. Among all TTS models, we can see that the MAE of the energy for CAI-TTS is smaller than all baselines.

% \vspace{-2mm}
% \subsubsection{Duration}
% \vspace{-2mm}

\noindent\textbf{Duration}: We also evaluate the quality of phoneme duration by calculating the distance~\cite{ren2020fastspeech} between the predicted duration and the ground-truth duration at a phoneme level.  
Note that the accented TTS corpus have provided the phoneme level duration.
For speech generated by Tacotron2 and Transformer TTS, we extract the phoneme duration from the trained attention alignment. For speech generated by FastSpeech2 and CAI-TTS, we obtain the duration from duration predictor output directly.
The seventh column of Table \ref{tab:tab2} shows that CAI-TTS generates more accurate phoneme duration than all benchmarking baselines.

\begin{figure}[t]
    \centering
    % \vspace{-4mm}
    \centerline{
    \includegraphics[width=0.435\linewidth]{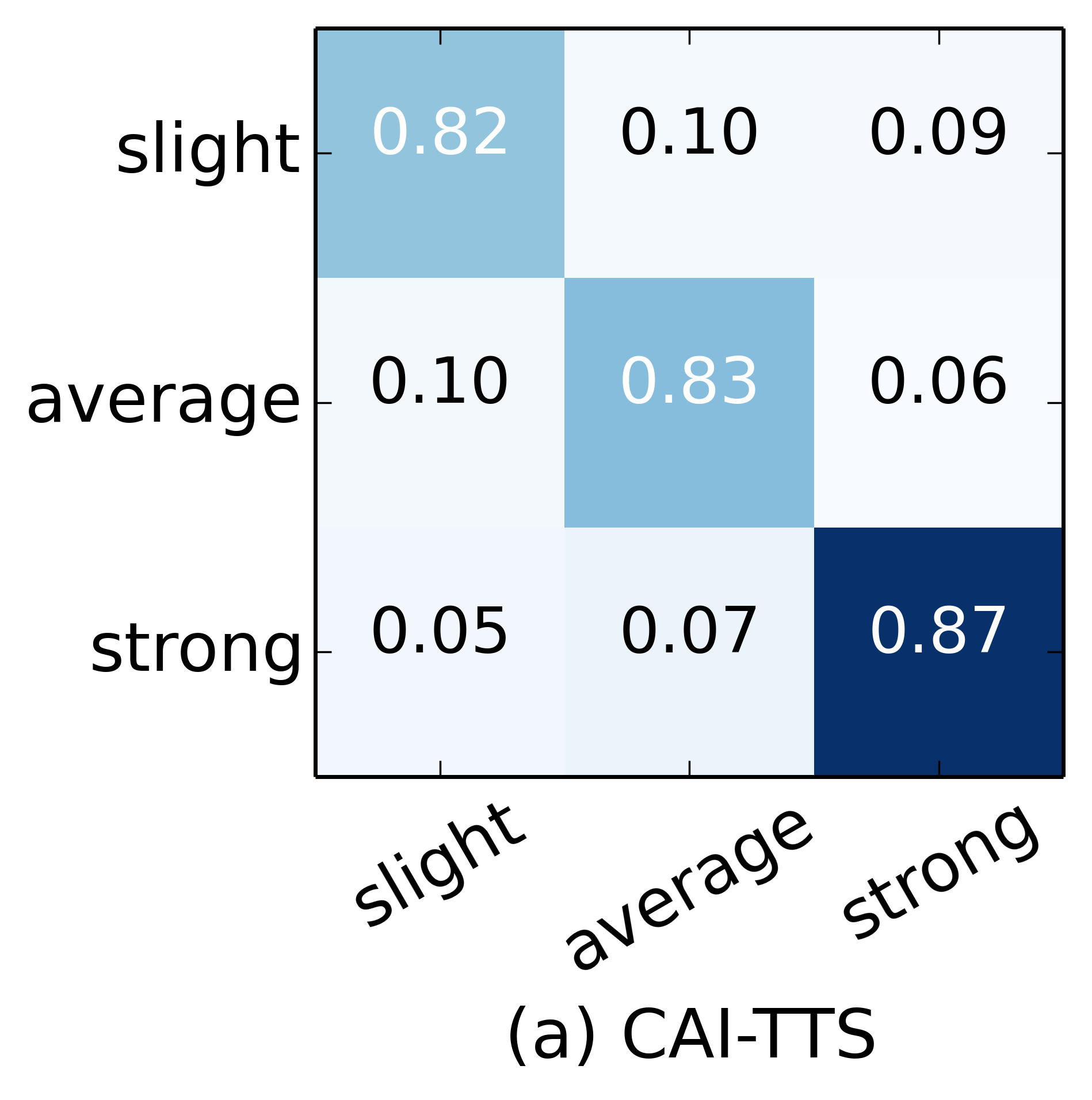}
    \includegraphics[width=0.54\linewidth]{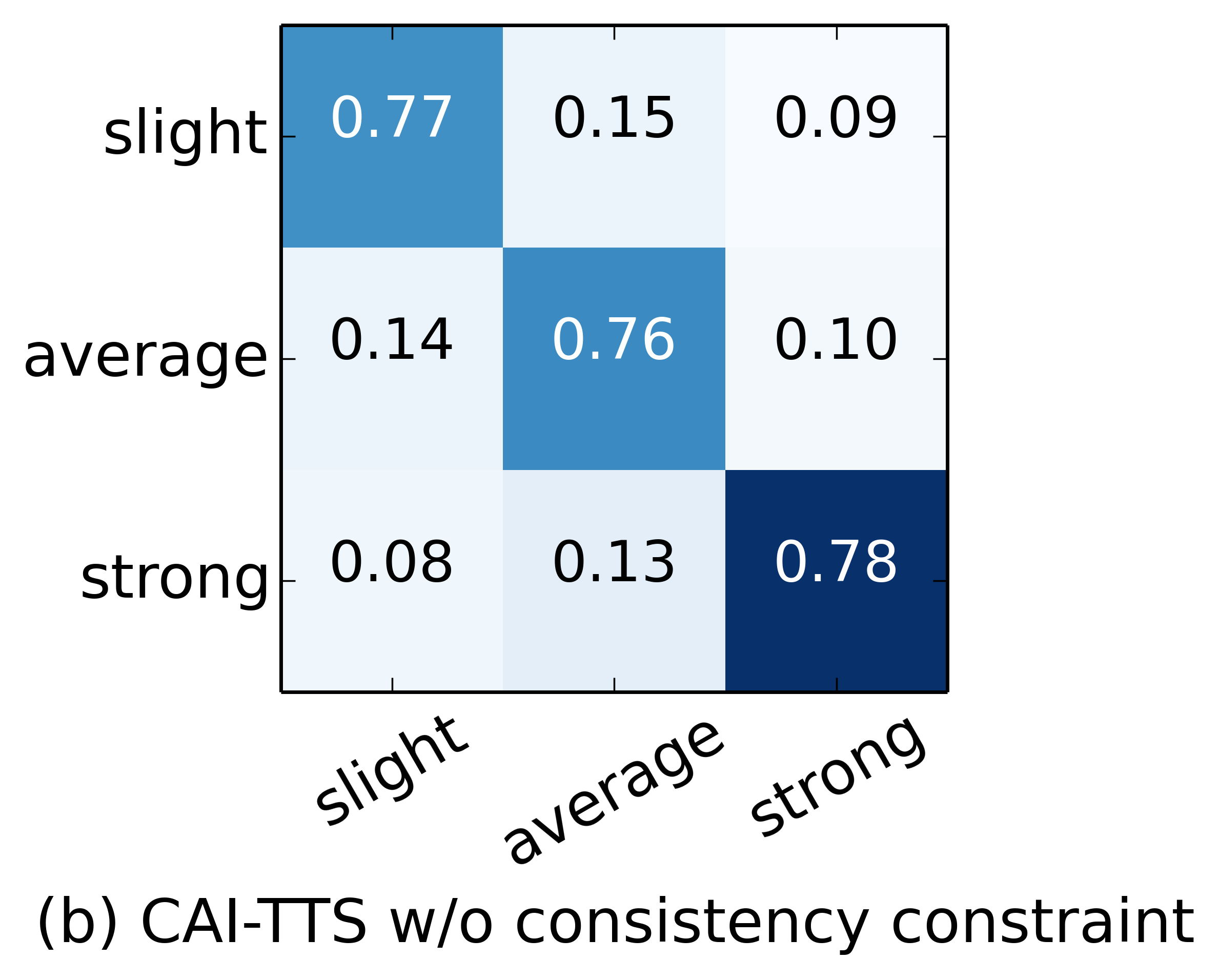}
    }
 
    \caption{Confusion matrices between perceived and intended accent intensity categories of synthesized speech.  (a) CAI-TTS;  b) CAI-TTS w/o consistency constraint.  The X-axis and Y-axis of the figures represent the perceived and intended category, namely slight, average, and strong.  }
    \label{fig:matirx1}
 
\end{figure}

\begin{figure}[!t]
    \centering
    % \vspace{-21mm}
   \centerline{
    \includegraphics[width=0.72\linewidth]{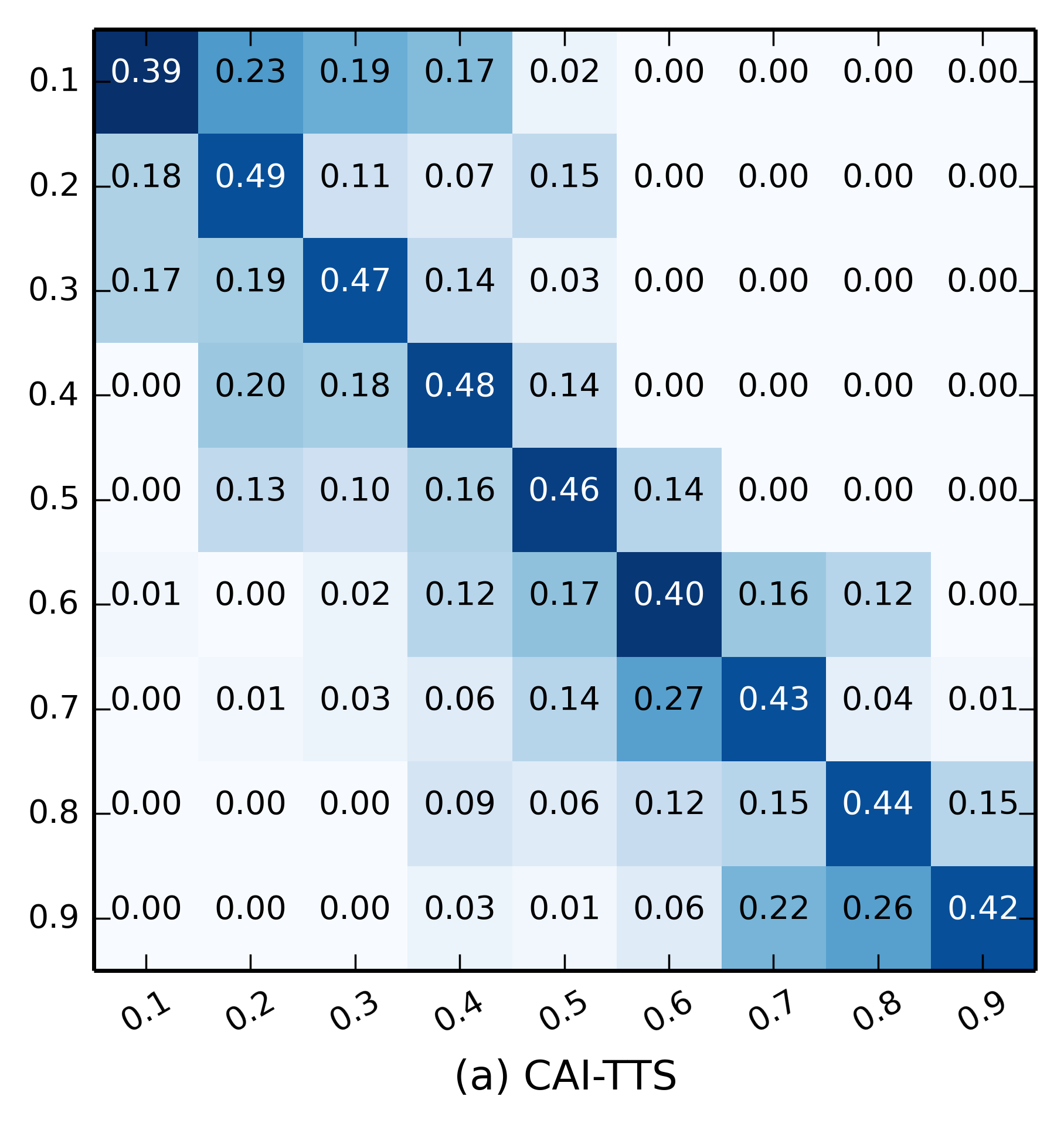}
    }
    \centerline{
    \includegraphics[width=0.72\linewidth]{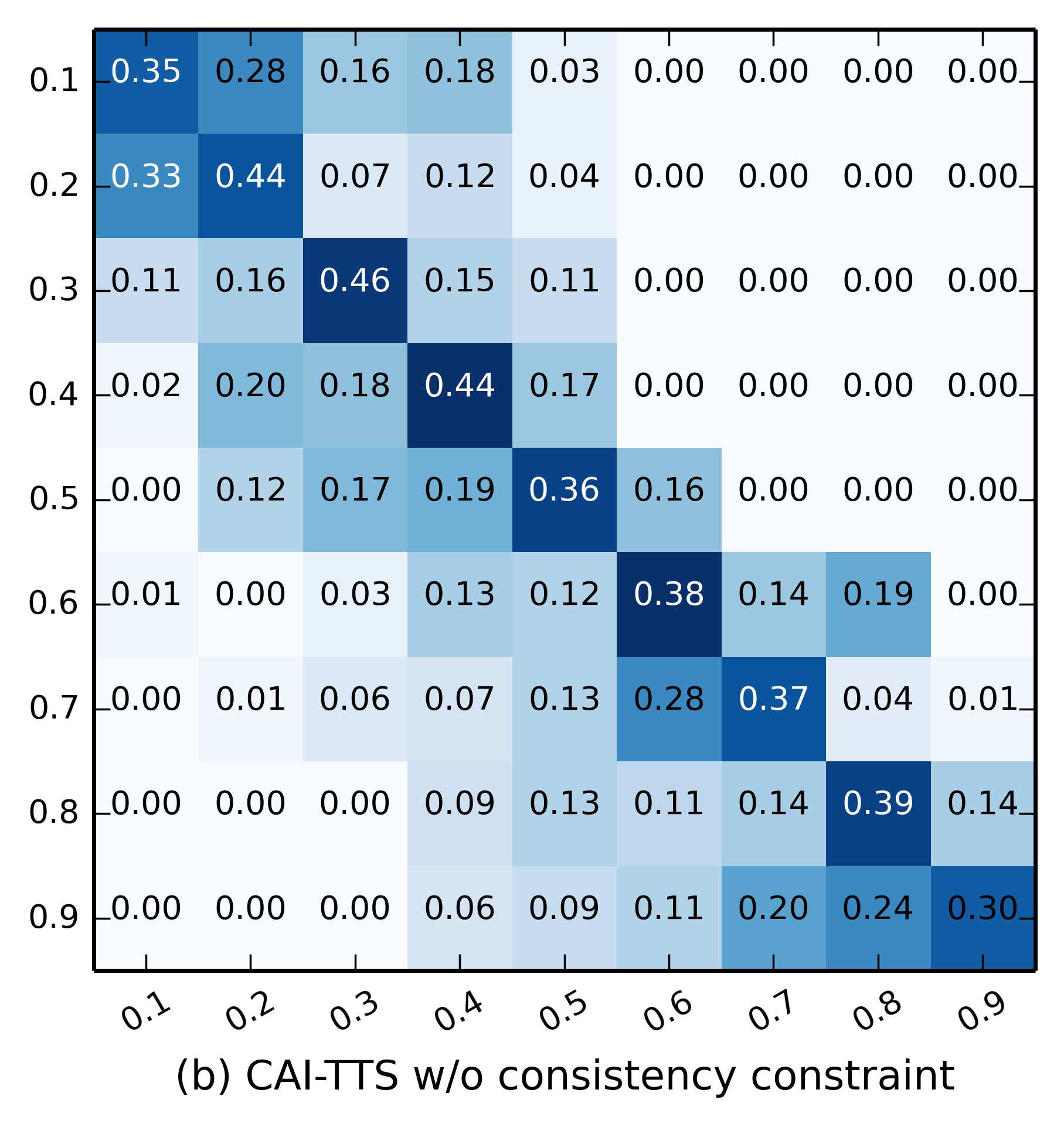}
    }
 
    \caption{Confusion matrices of synthesized speech from the a) CAI-TTS and b) CAI-TTS w/o consistency constraint. The X-axis and Y-axis of subfigures represent perceived and truth accent intensity scalar (ie. 0.1 to 0.9), respectively.}
    \label{fig:matirx2}
 
\end{figure}

%All observations for pitch, energy and duration indicate that our proposed CAI-TTS can indeed synthesize speech with more natural L2-accented pitch, energy and duration, which can result in better performance in terms of accentedness.

\subsection{Controllable Accent Intensity}
 
We further evaluate the ability of CAI-TTS to adjust the L2 accent intensity of the synthesized accented speech by comparing CAI-TTS with the CAI-TTS w/o consistency constraint. %, since there are no other relevant literature reporting a similar fine-grained control function for L2 accent.

We first conduct an intensity classification experiment.
At run-time, we assign the L2 speaker ID, accent ID and its intensity from 0.1 to 0.9 to synthesize the L2-accented speech with various accents. We consider the intensity scalars from 0.1 to 0.3 as `slight', 0.4 to 0.6 as `average' and 0.7 to 0.9 as `strong' in three categories.
We select 100 utterances from the test set, resulting in 100 samples for both systems.
Accordingly, all listeners are instructed to rate the accent intensity category, that are `slight', `average' or `strong', for each sample.
A listener can listen to the samples multiple times when needed.

Fig. \ref{fig:matirx1} presents the intensity confusion matrices.
It is observed that the CAI-TTS system shows a higher correlation between the perceived and intended accent intensity categories, with a correlation of over 80\%, that is considered a competitive result against other intensity-controlled studies. 
Furthermore, the CAI-TTS system clearly outperforms the contestant. The experiments confirm the superiority of the proposed controllable intensity mechanism.

We further 
% conduct a listening experiment to 
evaluate the intensity-controlled speech at a fine level. For each utterance, nine speech samples with intensity from 0.1 to 0.9 produced by CAI-TTS and CAI-TTS w/o consistency constraint form two groups. We invite 20 listeners, each listening to all samples in an increasing order of accent intensity.
The confusion matrices are reported in Fig. \ref{fig:matirx2}. 
We observe that the proposed CAI-TTS system provides a higher correlation between the perceived and the intended fine categories than the contestant.  
All subjective evaluations show consistently results that the CAI-TTS system provides an effective fine level intensity control.
% We would like to invite the readers to visit https://speechdemo.github.io/caitts for listening to the synthesized speech by the proposed CAI-TTS system.

\section{Conclusion}
\label{sec:con}
 
We have studied a novel TTS model, named CAI-TTS, to control the L2 accent and its intensity during the speech generation. 
We have conducted a series of experiments on audio quality, accent variance and intensity control to validate the effectiveness of the CAI-TTS model. The proposed CAI-TTS consistently outperforms all baselines in terms of accent rendering and the control of its intensity. This work marks an important step towards controllable rendering of accented TTS synthesis. 
%As our CAI-TTS controls the accent intensity at utterance level, hence in the future, we would like to study fine-grained (eg.  phoneme level) or hierarchical control methods. 
For future work, we plan to extend CAI-TTS to support fine-grained (eg. phoneme level) accent control.

\normalem
\bibliographystyle{IEEEtran}
{\footnotesize
\bibliography{refs}}

\begin{IEEEbiography}[{\includegraphics[width=1in,height=1.25in,clip,keepaspectratio]{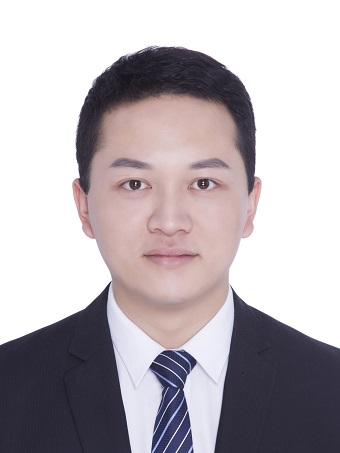}}]{Rui Liu}
is currently a Professor in National and Local Joint Engineering Research Center of Mongolian Intelligent Information Processing, Inner Mongolia University. Rui Liu received Ph.D degree from Inner Mongolia University, China in 2020 and Bachelor degree in Taiyuan University of Technology, ShanXi, China in 2014. From 2019 to 2020, he has been an exchange PhD candidate at the Department of Electrical \& Computer Engineering of National University of Singapore (NUS), funded by China Scholarship Council (CSC). From 2020 to 2022, he worked as a Research Fellow at the Department of Electrical and Computer Engineering, National University of Singapore, Singapore. He was the recipient of the ``Best Paper Award'' at the 2021 International Conference on Asian Language Processing (IALP).
He has published more than 20 papers in top-tier NLP/ML/AI conferences and journals, including IEEE/ACM-TASLP, Neural Networks, ICASSP, COLING, INTERSPEECH, etc. 
% Dr. Liu serves as the reviewer for many major referred journal and conference papers.
He is a member of IEEE, ISCA, CAAI and CCF, and serves as the reviewer for many major referred journal and conference papers.
His research interests broadly lie in audio, speech and natural language processing, which include expressive Text-to-Speech (TTS), expressive voice conversion, speech emotion recognition, prosody structure prediction, grapheme-to-phoneme conversion (G2P), syntax parsing et. al.
% His research interests include prosody and acoustic modeling for speech synthesis, voice conversion, machine learning and natural language processing.
% He has published more than 15 publications in major journals and international conferences, such as IEEE-T-ASLP, Neural Networks, IEEE-SPL, ICASSP, INTERSPEECH, \textit{etc.}
\end{IEEEbiography}

%  \newpage

\begin{IEEEbiography}[{\includegraphics[width=1in,height=1.25in,clip,keepaspectratio]{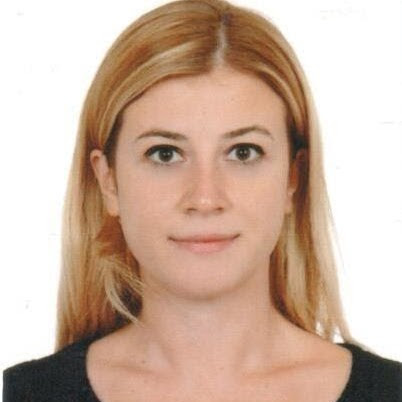}}]{Berrak Sisman}
received the PhD degree in Electrical and Computer Engineering from National University of Singapore in 2020, fully funded by A*STAR Graduate Academy under Singapore International Graduate Award (SINGA). She is currently working as a tenure-track Assistant Professor at the Erik Jonsson School Department of Electrical and Computer Engineering at University of Texas at Dallas, United States. Prior to joining UT Dallas, she was a tenure-track faculty at Singapore University of Technology and Design (2020-2022). She was a Postdoctoral Research Fellow at the National University of Singapore (2019-2020), and a Visiting Researcher at Columbia University, New York, United States (2020). She was an exchange PhD student at the University of Edinburgh and a visiting scholar at The Centre for Speech Technology Research (CSTR), University of Edinburgh in 2019. She was attached to RIKEN Advanced Intelligence Project, Japan in 2018. Her research is focused on machine learning, signal processing, emotion, speech synthesis and voice conversion. She plays leadership roles in conference organizations and also active in technical committees. She has served as the General Coordinator of the Student Advisory Committee (SAC) of International Speech Communication Association (ISCA). She has served as the Area
Chair at INTERSPEECH 2021, INTERSPEECH 2022, IEEE SLT 2022 and as the Publication Chair at ICASSP 2022. She has been elected
as a member of the IEEE Speech and Language Processing Technical Committee (SLTC) in the area of Speech Synthesis for the term from Jan. 2022 to Dec. 2024.
\end{IEEEbiography}

% \newpage
\begin{IEEEbiography}[{\includegraphics[width=1in,height=1.25in,clip,keepaspectratio]{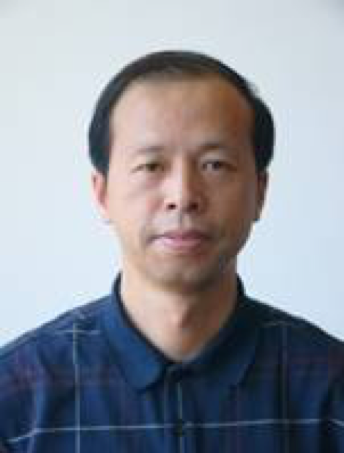}}]{Guanglai Gao}
received the B.S. degree from Inner Mongolia University, Hohhot, China, in 1985, and the
M.S. degree from the National University of Defense Technology, Changsha, China, in 1988.
He was a Visiting Researcher at University of Montreal, Canada.
Currently, he is a Professor with the Department of Computer Science, Inner Mongolia University. His research interests include artificial intelligence and pattern recognition.
\end{IEEEbiography}
 
 \vspace{10mm}
\begin{IEEEbiography}[{\includegraphics[width=1in,height=1.25in,clip,keepaspectratio]{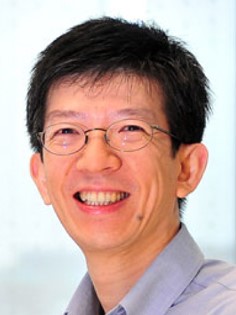}}]{Haizhou Li}
Haizhou Li (M’91-SM’01-F’14) received the B.Sc., M.Sc., and Ph.D degree in electrical and electronic engineering from South China University of Technology, Guangzhou, China in 1984, 1987, and 1990 respectively. Dr Li is currently a Professor at the School of Data Science, the Chinese University of Hong Kong, Shenzhen, China, and the Department of Electrical and Computer Engineering, National University of Singapore (NUS). His research interests include automatic speech recognition, speaker and language recognition, and natural language processing. Prior to joining NUS, he taught in the University of Hong Kong (1988-1990) and South China University of Technology (1990-1994). He was a Visiting Professor at CRIN in France (1994-1995), Research Manager at the Apple-ISS Research Centre (1996-1998), Research Director in Lernout \& Hauspie Asia Pacific (1999-2001), Vice President in InfoTalk Corp. Ltd. (2001-2003), and the Principal Scientist and Department Head of Human Language Technology in the Institute for Infocomm Research, Singapore (2003-2016). Dr Li served as the Editor-in-Chief of IEEE/ACM Transactions on Audio, Speech and Language Processing (2015-2018), a Member of the Editorial Board of Computer Speech and Language (2012-2018). He was an elected Member of IEEE Speech and Language Processing Technical Committee (2013-2015), the President of the International Speech Communication Association (2015-2017), the President of Asia Pacific Signal and Information Processing Association (2015-2016), and the President of Asian Federation of Natural Language Processing (2017-2018). He was the General Chair of ACL 2012, INTERSPEECH 2014 and ASRU 2019. Dr Li is a Fellow of the IEEE and the ISCA. He was a recipient of the National Infocomm Award 2002 and the President’s Technology Award 2013 in Singapore. He was named one of the two Nokia Visiting Professors in 2009 by the Nokia Foundation, U Bremen Excellence Chair Professor in 2019, and Fellow of Academy of Engineering Singapore in 2022.

\end{IEEEbiography}
% \end{multicols}
% \enlargethispage{-9.5cm}
% \enlargethispage{2mm}
\end{document}